\documentclass[english]{article}
\usepackage[latin9]{inputenc}
\usepackage{amsmath}
\usepackage{graphicx}
\makeatletter
\providecommand{\tabularnewline}{\\}
\usepackage{babel}
\usepackage{subfig}
\usepackage{caption}
\usepackage{stackengine}
\usepackage{float}
\makeatother

\begin{document}

\title{Relation between exponential behavior and energy denominators-Weak
Coupling Limit}

\author{Levering Wolfe and Larry Zamick \footnote{lzamick@physics.rutgers.edu} \\
 Department of Physics and Astronomy\\
 Rutgers University, Piscataway, New Jersey 08854 }
\maketitle
\begin{abstract}
We show some interesting properties of tridiagonal and pentadiagonal
matrices in the weak coupling limits. In the former case of this limit
the ground state wave function amplitudes are identical to the Taylor
expansion coefficients of the exponential function e$^{(-\frac{v}{e})}$.
With regards to transition rates a dip in the pentadiagonal case which
is not present in the tridiagonal case is explained. An intimate connection
between energy denominators and exponential behavior is demonstrated.

\textit{Keywords:} Distribution ,\\
 $ $\\
 PACs Number: 21.60.Cs 
\end{abstract}

\section{Introduction}

This work is a continuation of work done before on matrix models of
strength distributions{[}1,2{]}.Matrix mechanics was of course introduced
into quantum mechanics by Heisenberg{[}2{]} and Born and Jordan{[}3{]}.
In the previous work we had a matrix in which the diagonal elements
were E$_{n}$= nE with E=1 MeV. We introduced a constant coupling v which
for a level E$_{n}$ occurs only with the nearest neighbors E$_{(n-1)}$
and E$_{(n+1)}.$ The matrix is shown in Table 1. Note that the only
relevant parameter is $\frac{v}{E}$. Here we extend the work to pentadiagonal
matrices, as shown in Table 2. In both cases we discuss for the first time the weak coupling
limits for the ground state wave functions.

\vspace{150pt}
\begin{minipage}[t]{1\columnwidth}%
Table 1: The Matrix Hamiltonian-Tridiagonal%
\end{minipage}

\begin{tabular}{|c|c|c|c|c|c|c|c|c|c|c|}

\hline 
0  & v  & 0  & 0  & 0  & 0  & 0  & 0  & 0  & 0  & 0\tabularnewline
\hline 
v  & E  & v  & 0  & 0  & 0  & 0  & 0  & 0  & 0  & 0\tabularnewline
\hline 
0  & v  & 2E  & v  & 0  & 0  & 0  & 0  & 0  & 0  & 0\tabularnewline
\hline 
0  & 0  & v  & 3E  & v  & 0  & 0  & 0  & 0  & 0  & 0\tabularnewline
\hline 
0  & 0  & 0  & v  & 4E  & v  & 0  & 0  & 0  & 0  & 0\tabularnewline
\hline 
0  & 0  & 0  & 0  & v  & 5E  & v  & 0  & 0  & 0  & 0\tabularnewline
\hline 
0  & 0  & 0  & 0  & 0  & v  & 6E  & v  & 0  & 0  & 0\tabularnewline
\hline 
0  & 0  & 0  & 0  & 0  & 0  & v  & 7E  & v  & 0  & 0\tabularnewline
\hline 
0  & 0  & 0  & 0  & 0  & 0  & 0  & v  & 8E  & v  & 0\tabularnewline
\hline 
0  & 0  & 0  & 0  & 0  & 0  & 0  & 0  & v  & 9E  & v\tabularnewline
\hline 
0  & 0  & 0  & 0  & 0  & 0  & 0  & 0  & 0  & v  & 10E\tabularnewline
\hline 
\end{tabular}\vspace{20pt}

Table 2: The Matrix Hamiltonian-Pentadiagonal

\begin{tabular}{|c|c|c|c|c|c|c|c|c|c|c|}

\hline 
0  & v  & v  & 0  & 0  & 0  & 0  & 0  & 0  & 0  & 0\tabularnewline
\hline 
v  & E  & v  & v  & 0  & 0  & 0  & 0  & 0  & 0  & 0\tabularnewline
\hline 
v  & v  & 2E  & v  & v  & 0  & 0  & 0  & 0  & 0  & 0\tabularnewline
\hline 
0  & v  & v  & 3E  & v  & v  & 0  & 0  & 0  & 0  & 0\tabularnewline
\hline 
0  & 0  & v  & v  & 4E  & v  & v  & 0  & 0  & 0  & 0\tabularnewline
\hline 
0  & 0  & 0  & v  & v  & 5E  & v  & v  & 0  & 0  & 0\tabularnewline
\hline 
0  & 0  & 0  & 0  & v  & v  & 6E  & v  & v  & 0  & 0\tabularnewline
\hline 
0  & 0  & 0  & 0  & 0  & v  & v  & 7E  & v  & v  & 0\tabularnewline
\hline 
0  & 0  & 0  & 0  & 0  & 0  & v  & v  & 8E  & v  & v\tabularnewline
\hline 
0  & 0  & 0  & 0  & 0  & 0  & 0  & v  & v  & 9E  & v\tabularnewline
\hline 
0  & 0  & 0  & 0  & 0  & 0  & 0  & 0  & v  & v  & 10E\tabularnewline
\hline 
\end{tabular}

.

\section{The calculation}

The eigenfunctions resulting from the diagonalizations of the above
matrices are of the form

\{a$_{0}$,a$_{1}$,a$_{2,}$, .....a$_{10}$\}. We are interested
in the values of these a$_{n}$ in the limit where $\frac{v}{E}$ is very small
(NB In previous publications we used the notation a$_{1}$ to a$_{11}$).

We ran Mathematica programs for small v to get an idea of the coefficients
of the ground state eigenfunctions. For the tridiagonal case we used
$\frac{v}{E}$=0.01 . The results were as follows:

\{0.99995, --0.009999, 0.0000499933, -1.6664{*}10$^{-7}$, 4.16592{*}
10$^{-10}$, -8.33171{*} 10$^{-13}$, 1.3886{*} 10$^{-15}$, -1.98269{*}
10$^{-18}$, 2.47958{*} 10$^{-21}, -2.75506${*} 10$^{-24}$, 2.75504{*}
10$^{-27}$\}

These numbers can be put in a more suggestive way with a bit of rounding
up.

1, -$\frac{v}{E}$, ($\frac{v^2}{E^22}$), -($\frac{v^3}{E^3 6}$) ......... (-1)$^{n}$ $\frac{v^n}{E^nn!}$.
We recognize these a$_{i}$'s as coefficients in the Taylor series
of e$^{-v}$$^{/E}$.

To derive this result we should realize that to get a$_{n}$ we have
to go the nth order in perturbation theory. Let H=H$_{0}$+V with
H$_{0}$ the diagonal part of the matrix and V the off diagonal.

$$\Psi= \varPhi + \frac{1}{E_{0}-H_{0}} QV \varPsi= (1+\frac{1}{E_{0}-H_{0}}QV+\frac{1}{E_{0}-H_{0}}\hspace{5pt}QV \frac{1}{E_{0}-H_{0}}\hspace{5pt}QV+..... )\varPhi$$

where Q prevents the unperturbed ground state from being an intermediate
state.

To get a$_{n}$ to the lowest power in $\frac{v}{E}$ we have to go in n steps
0 to 1, 1 to 2,...n-1 to n. In the numerator all the matrix elements
\textless{} (n+1) QVn\textgreater{} are the same, namely v. In
the denominator we get (E$_{0}$-E$_{1})$.....(E$_{0}$ -E$_{n}$).
Since we have E$_{n}$ = nE the denominator is $(n!)$E$^n$. This proves the
result and shows an intimate connection between exponential behavior
and equally spaced levels.

We next consider the pentadiagonal case. This time we chose $\frac{v}{E}$ to
be 10$^{-10}$. The results are are as follows:

\{ \{1, - 1.00006{*}10$^{-10}$, - 5.0{*}10$^{-11}$, 5.00091{*}10$^{-21}$,
1.25{*}10$^{-21}$, - 1.25071{*}10$^{-31}$, - 2.08333{*}10$^{-32}$,
 2.08951{*}10$^{-42}$, 2.60417{*}10$^{-43}$, - 2.63376{*}10$^{-53}$, -
2.60417{*}10$^{-54}$\} \},

These numbers can be put in a more suggestive way with a bit of rounding
up.

\{1,- $\frac{v}{E}$, -$\frac{v}{2E}$, ($\frac{v^2}{E^2 2}$), ($\frac{v^2}{E^2 8}$), -( $\frac{v^3}{E^3 8}$)
.....

We now have 2 types of non-vanishing matrix elements -one from m to(m+1)
and another from m to (m+2). Both have a value v so it is still the
energy denominators which come into play. We now consider selected
a$_{n}$:

a$_{1}$: We get $\frac{v}{E_{0}-E_{1}}$ =- $\frac{v}{E}$.

a$_{2}$: In order to get a result linear in v we have only the direct
connection from 0 to 2 which yields a value $-\frac{v}{2} $

a$_{3}$: there are 2 paths:

A: 0 to 2 and 2 to 3

B: 0 to 1 and 1 to 3.

In the former case the value is $\frac{1}{2*3} $ and in the latter $\frac{1}{3}$.
So we get $\frac{1}{6}+\frac{1}{3}=\frac{1}{2}$--answer $-\frac{v}{2E}$.

a$_{4}$: We look for($\frac{v}{E}$)$^{2}$ terms. There is actually only one:
0 to 2 and 2 to 4. This gives a value $\frac{1}{2*4}$=$\frac{1}{8} $-answer $\frac{1}{8}$ ($\frac{v}{E}$)$^{2}$

a$_{5}$: There are 3 paths:

C: 0 to 1, 1 to 3, 3 to five: $-\frac{1}{1*3*5} $

D: 0 to 2, 2 to 3, 3 to 5: $-\frac{1}{2*3*5}$

E: 0 to 2, 2 to 4, 4 to 5: $-\frac{1}{2*4*5}$ 

sum = $-\frac{1}{8}$ - answer $-\frac{1}{8}$ ($\frac{v}{E}$)$^{3}$

And so it goes.

\section{Results for \textless{}n T$_{1}$(n+1)\textgreater{} and \textless{}n
T$_{2}$ (n+1)\textgreater{} }

In previous works we introduced 2 possible transition operators:

\textless{}n T$_{1}$(n+1)\textgreater{} is a constant. We choose
this constant to be one.

\textless{}(n+1)T$_{2}$n\textgreater is equal to $\sqrt{n+1}$

In both cases all other transition elements are taken to be zero. Since this is a model we do not have to assign units to T$_1$ and T$_2$, but it should be pointed out that if we were to say double T$_1$ we would quadruple the transition rate. This would affect the scale but not the overall shape of the graph.   

The strength matrix element between a state \{a\} and a state \{b\}
for T$_{1}$ is simply 
$$ O=(a_{0}b_{1}+,..+a_{9} b_{10}) + (b_{0}a_{1}+....+b_{9}
a_{10})\textless{}n T_{1}(n+1)\textgreater{} $$
with the last factor taken to be unity (1). For T$_{2}$ we take the previous expression and multiply the a$_{n}$ b$_{(n+1)}$ by $\sqrt{n+1}$ -likewise
b$_{n}$ a$_{(n+1)}$. The strength is O$^2$ and we will be plotting ln(O$^2$) versus excitation energy  E$^*$.

In Fig.1 and 2 we show results for T$_{1}$, v=0.1 MeV for the tridiagonal
and pentadiagonal cases. In Fig 3 and 4 we show corresponding results
for T$_{2}$. We see a near exponential decrease of the strength in
the the tridiagonal case with T$_{1}$. This shows up on our log plot
as a near straight line with a negative slope, however the last point
(the tenth excited state) is much lower than the exponential projection.
This has been discussed in {[}1,2{]} where it was shown that for $T_{1}$
the value is actually minus infinity. Also it should be pointed out
that although for v=0.1 MeV the tridiagonal curve for T$_{1}$ looks exponential
and this persists until v=1 and even beyond, this is not the case
for extremely small v or extremely large v. This can be seen in the
figures of ref{[}2{]}. 

In Table 3 we show what happens for very small v. In the weak coupling limit the T$_1$ matrix
element O can be shown to be of the form O(1$\rightarrow$n)= $(\frac{v}{E})^m$A(1$\rightarrow$n) where A is independent of $\frac{v}{E}$. We define g$_k$=$\frac{1}{k!}$ and give the values of A in the third column of the table. In the fourth column of Table 3 we give the value of ln(O$^2$) using the expressions in the second and third columns
i.e. keeping only the lowest powers of $\frac{v}{E}$. We do this for $\frac{v}{E}$= 0.0001. The behavior is
somewhat complicated.The results are not monotonic in n. For example although m increases from n=1 to 5, there is a decease in going from n=5 (m=6) to n=6 (m=5). Thus we get a non-exponential behavior with the value of ln(O$^2$ ) being larger for n=6 than for n=5. However when we go to larger $\frac{v}{E}$ e.g. v= 0.1, the behavior looks quite exponential.  For T$_2$, on the other hand ,we do get an exponential  behavior in the  extreme weak coupling limit i.e. m=n for all n.
\\

\begin{minipage}[t]{1\columnwidth}%
Table 3: Weak coupling expression for The T$_{1}$ amplitude and ln
(O$^{2}$) .%
\end{minipage}

\begin{tabular}{|c|c|c|c|}
\hline 
n & m & A(1$\rightarrow n$) & ln(O$^{2})[v=0.0001${]}\tabularnewline
\hline 
1 & 0 & 1 & 0\tabularnewline
\hline 
2 & 3 & -g$_2$-g$_3$ & -56.07\tabularnewline
\hline 
3 & 4 & g$_2^2$-g$_3$+g$_4$ & -77.84\tabularnewline
\hline 
4 & 5 & -g$_4$-g$_5$ & -96.09\tabularnewline
\hline 
5 & 6 & g$_4$-g$_3^2$+g$_6$ & -118.89\tabularnewline
\hline 
6 & 5 & -g$_5$ & -101.67\tabularnewline
\hline 
7 & 6 & g$_4$-g$_3^2$-g$_5$+g$_6$ & -120.43\tabularnewline
\hline 
8 & 7 & -g$_5$*g$_2$+g$_6$-g$_7$ & -140.46\tabularnewline
\hline 
9 & 8 & g$_4^2$-g$_5$*g$_3$+g$_6$*g$_2$-g$_7$+g$_8$ & -161.46\tabularnewline
\hline 
10 &  & 0 & minus infinity\tabularnewline
\hline 
\end{tabular}
\\

For very large v there is an even-odd effect resulting in 2 exponential behaviors-one for even n and one or odd n. In Fig. 2 we show corresponding results for the pentadiagonal case.
There is one significant difference between tri and penta for T$_{1}$.
As seen in Fig 2 in the transition from the ground state to the second
excited state there is a big dip for the pentadiagonal case. This
is not the case for the tridiagonal case. After the dip in the pentadiagonal
case one gets a near exponential behavior.
\vspace{20pt}

\vspace{10pt}
\begin{center}

\begin{figure}[h]%
\footnotesize
\stackunder[5pt]{\includegraphics[width=5cm,height=3cm]{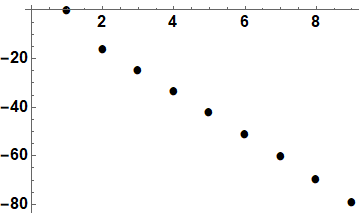}}{Figure 1: T$_1$ Tridiagonal v=0.1}%
\hspace{1cm}%
\stackunder[5pt]{\includegraphics[width=5cm,height=3cm]{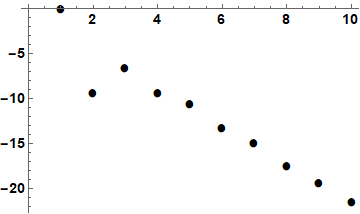}}{Figure 2: T$_1$ Pentadiagonal  v=0.1}
\caption*{}
\end{figure}

\end{center}

\vspace{10pt}

To explain the dip we examine the second excited state for the pentadiagonal
case in the weak coupling limit. Recall that for the ground state
the values of \{ a$_{0}$, a$_{1}$, a$_{2}$\} are respectively \{1,-$\frac{v}{E}$,$-\frac{v}{2E}$\}.
For the second excited state the values of \{b$_{0}$,b$_{1}$,b$_{2}$\}
can be shown to be \{$\frac{v}{2E}$, v, 1\}. For the T$_{1}$ case note that
a$_{0}$*b$_{1}$=$\frac{v}{E}$ while a$_{1}$*b$_{2}$ = -$\frac{v}{E}$ so we have complete
cancellation of the leading terms.

To get \{b$_{0}$,b$_{1}$,b$_{2}$\} we note that clearly b$_{2}$=1
in the weak coupling limit. For b$_{0}$ we go from state 2 to state
0 so that we get a contribution (from the energy denominator) $\frac{v}{2E}$
i.e. ( E$_{2}$-E$_{0}$)=2. For b1 we go from state 2 to state 1
and so we get $\frac{v}{E}$ i.e. (E2-E1)=1.

For the T$_{_{2}}$ case , figs 3 and 4, the dip is not as pronounced.
Because of the factor $\sqrt{(n+1)}$ there is only a partial cancellation.

\begin{figure}[h]%
\footnotesize
\stackunder[5pt]{\includegraphics[width=5cm,height=3cm]{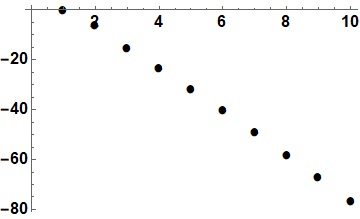}}{Figure 3:  T$_2$ Tridiagonal v=0.1 }%
\hspace{1cm}%
\stackunder[5pt]{\includegraphics[width=5cm,height=3cm]{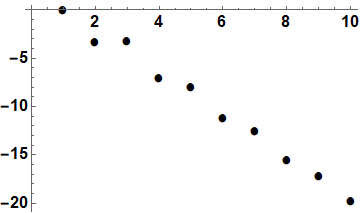}}{Figure 4:  T$_2$ Pentadiagonal  v=0.1}
\caption*{}
\end{figure}

\vspace{10pt}
 We had previously discussed the strong coupling limit for T$_{1}$ {[}1,2{]}.
In that case all transitions vanished. This was explained by the fact
that in the strong coupling limit the transition matrix becomes identical
to the Hamiltonian.

\section{All v's.}

We briefly consider the case where on the diagonal we still have nE
but all other matrix elements are v (there are no zeros). One can easily
show that the ground state column vector \{a$_{0}$, a$_{1}$,....a$_{10)}$\}
in the weak coupling limit is \{1,-$\frac{v}{E}$,$-\frac{v}{2E}$,$-\frac{v}{3E}$ ,...$-\frac{v}{10E}$ 
\}.

There are other matrix models which address problems related to, but
different, from what we have here considered. As previously mentioned,
{[}1{]} Bohr and Mottelson {[}5{]} used matrix models to derive the
Breit-Wigner formula for a resonance. Brown and Bolsterli described
the giant dipole resonances in nuclei in a schematic model using a
delta interaction{[}6{]}. In that work they made the approximation
that certain radial integrals were constant. Abbas and Zamick {[}7{]}
removed this restriction. Generally speaking matrix models are very
useful for casting insights into the physics of given problems where
the more accurate but involved calculations fail.

Although the models here are not geared to fit specific experiments
we do keep our eyes on the empirical data. In refs {[}8-11{]} we cite
works in which exponential behavior is seen and explained. Here we
show that with unperturbed equally spaced energy levels we generate
factorials which are needed to get exponential behavior. These come
from the energy denominators. Also, our perturbations are simple enough
that we can easily perform nth order perturbation theory for any n
and get analytic results in the weak coupling limit. 
\vspace{15pt}

\vspace{100pt}

\section{Appendix}

\vspace{10pt}

\begin{table}[H]

\caption*{Table 4: Wave functions for tridiagonal case in the weak coupling limit. Note that u = $\frac{v}{E}$}
\begin{tabular}{|c|c|c|c|c|c|c|c|c|c|c|}
\hline 
a$_{0}$ & a$_{1}$ & a$_{2}$ & a$_{3}$ & a$_{4}$ & a$_{5}$ & a$_{6}$ & a$_{7}$ & a$_{8}$ & a$_{9}$ & a$_{10}$\tabularnewline
\hline 
1 & -u & u$^{2}$/2! & -u$^{3}/3!$ & u$^{4}$/4! & -u$^{5}$/5! & u$^{6}$/6! & -u$^{7}$/7! & u$^{8}$/8! & -u$^{9}$/9! & u$^{10}$/10!\tabularnewline
\hline 
u & 1 & -u & u$^{2}$/2! & -u$^{3}/3!$ & 0 & 0 & 0 & 0 & 0 & 0\tabularnewline
\hline 
u$^{2}$/2! & u & 1 & -u & u$^{2}$/2! & - u$^{3}$/3! & 0 & 0 & 0 & 0 & 0\tabularnewline
\hline 
u$^{3}$/3! & u$^{2}$/2! & u & 1 & -u & u$^{2}$/2! & - u$^{3}$/3! & 0 & 0 & 0 & 0\tabularnewline
\hline 
u$^{4}$/4! & u$^{3}$/3! & u$^{2}$/2! & u & 1 & -u & u$^{2}$/2! & - u$^{3}$/3! & 0 & 0 & 0\tabularnewline
\hline 
0 & u$^{4}$/4! & u$^{3}$/3! & u$^{2}$/2! & u & 1 & -u & u$^{2}$/2! & - u$^{3}$/3! & 0 & 0\tabularnewline
\hline 
0 & 0 & u$^{4}$/4! & u$^{3}$/3! & u$^{2}$/2! & u & 1 & -u & u$^{2}$/2! & - u$^{3}$/3! & 0\tabularnewline
\hline 
0 & 0 & 0 & u$^{4}$/4! & u$^{3}$/3! & u$^{2}$/2! & u & 1 & -u & u$^{2}$/2! & - u$^{3}$/3!\tabularnewline
\hline 
0 & 0 & 0 & 0 & u$^{4}$/4! & u$^{3}$/3! & u$^{2}$/2! & u & 1 & -u & u$^{2}$/2!\tabularnewline
\hline 
0 & 0 & 0 & 0 & 0 & u$^{4}$/4! & u$^{3}$/3! & u$^{2}$/2! & u & 1 & -u\tabularnewline
\hline 
u$^{10}$/10! & u$^{9}$/9! & u$^{8}$/8! & u7/7! & u$^{6}$/6! & u$^{5}$/5! & u$^{4}$/4! & u$^{3}$/3 & u$^{2}$/2 & u & 1\tabularnewline
\hline
\end{tabular}

\end{table}

Levering Wolfe was supported by a Rutgers Aresty Research Assistant award and a Richard J. Plano award.

\end{document}